\documentclass[aps,twocolumn,preprintnumbers]{revtex4}

\usepackage{graphicx}  
\usepackage{subfigure}
\usepackage{multirow}

\usepackage{fancyhdr}
\usepackage{longtable}
\usepackage{parskip}
\usepackage[T1]{fontenc}
\usepackage{dcolumn}   
\usepackage[dvipsnames]{xcolor}
\usepackage{bm}        
\usepackage{amsfonts}  
\usepackage{amsmath}   
\usepackage{amssymb}   
\usepackage{hyperref}

\newcommand{\pwisein}{\left\{ \begin{array}{ll}}
\newcommand{\pwiseout}{\end{array}\right.}

\begin{document}

\title{Microwave dynamics of gated Al/InAs superconducting nanowires}

\author{Vittorio Buccheri$^1$\footnote{buccheri@chalmers.se}, Fran\c{c}ois Joint$^1$\footnote{Current address: Group for Advances Receiver Development (GARD), Department of Space, Earth and Environment, Chalmers University of Technology, 41296 Gothenburg, Sweden.}, Kazi Rafsamjani Amin$^1$, Tosson Elalaily$^{2,3,4}$, Olivér Kürtössy$^{2,3}$, Zoltán Scherübl$^{2,3}$, Gerg\H{o} Fülöp$^{2,3}$, Thomas Kanne$^5$, Jesper Nygård$^5$, Péter Makk$^{2,6}$, Szabolcs Csonka$^{2,3,7}$, Simone Gasparinetti$^1$\footnote{simoneg@chalmers.se}}

\affiliation {\it $^1$Department of Microtechnology and Nanoscience, Chalmers University of Technology, 41296 Gothemburg, Sweden.\\ $^2$Department of Physics, Institute of Physics, Budapest University of Technology and Economics, M\H{u}egyetem rkp. 3., H-1111 Budapest, Hungary.\\ $^3$MTA-BME Superconducting Nanoelectronics Momentum Research Group, M\H{u}egyetem rkp. 3., H-1111 Budapest, Hungary.\\ $^4$Low-Temperature Laboratory, Department of Applied Physics, Aalto University School of Science, P.O. Box 15100, FI-00076 Aalto, Finland. 
\\ $^5$Center for Quantum Devices, Niels Bohr Institute, University of Copenhagen, Universitetsparken 5, DK-2100 Copenhagen, Denmark.\\ $^6$MTA-BME Correlated van der Waals Structures Momentum Research Group, M\H{u}egyetem rkp. 3., H-1111 Budapest, Hungary.\\ $^7$Institute of Technical Physics and Materials Science, HUN-REN Centre for Energy Research, Konkoly-Thege Miklós út 29-33, H-1121 Budapest, Hungary.}

\begin{abstract}  

Several experiments have recently reported on gate-tunable superconducting properties in metallic devices, holding promise for the realization of cryogenic switches, tunable resonators, and superconducting logic. In particular, the suppression of the critical current as a function of the gate voltage has been widely investigated. However, time-domain studies are discussed only in a few cases. In this paper, we present a microwave characterization of a gate-controlled Al-capped InAs nanowire embedded in a $\lambda/4$ coplanar waveguide resonator. We observe a shift in the resonator frequency and an increase in its internal losses as a function of the gate voltage, which we relate to a change in the imaginary and real components of the nanowire impedance, respectively. We demonstrate that these changes are described by the Mattis-Bardeen model with an effective temperature. We further study the resonator response to fast-varying gate signals and measure characteristic response times of the order of 40 ns, both in time-domain and parametric modulation experiments. Our study elucidates the impact of the gate on the complex impedance of the nanowire in the superconducting state, as well as its dynamic performance, providing a foundation for the design of gate-controlled superconducting devices.

\end{abstract}

\maketitle 

Superconductors, thanks to their low energy losses and the Josephson effect, are widely used in cryoelectronics and quantum information processing \cite{Braginski,Kjaergaard}. In these applications, device-design considerations require a specific optimization of superconducting properties such as transition temperature, kinetic inductance, and critical current. For example,  kinetic inductance affects the resonant frequency of superconducting resonators \cite{Tinkham}, and the spacing of energy levels in transmon qubits depends on the critical current of Josephson junctions \cite{Devoret}. 

Although these properties are partially limited by material and geometry \cite{Tinkham}, there are strategies to tune them locally and in real time. In superconducting quantum interference devices, for example, a magnetic field is used to control the critical current in one or more Josephson junctions arranged in a superconducting loop \cite{Fagaly,Li}. On the other hand, hybrid superconducting-semiconducting systems take advantage of the low carrier density in the semiconducting channel to tune the critical current with an electric field \cite{Chen,Burkard}. 

Recently, several experiments \cite{DeSimoni,Paolucci2018,Paolucci2019,Puglia,Yu,Pashkin,RitterPhonon,Catto,Elalaily2021,Elalaily2023} have reported electrostatic gate control even in fully superconducting devices. The mechanism behind such findings, which was originally explained as a field effect \cite{DeSimoni, Paolucci2018, Paolucci2019}, is still under debate \cite{Pashkin,RitterPhonon,Catto,Elalaily2021,Elalaily2023,Elalaily2024,Ruf}. Nonetheless, gate-controlled superconducting devices promise the same flexibility of semiconductor-based technology in terms of local and real-time control, which would make them appealing for cryoelectronics applications such as superconducting switches \cite{RitterSwitch}, tunable resonators \cite{Ryu}, and fast superconducting electronics \cite{Elalaily2024,Joint}. A review summarizing the findings, outstanding questions, and possible application of this phenomenon is given in Ref.~\cite{Ruf_rev}.

An understanding of the dynamics and characteristic times of the gate-response in these devices is a key ingredient for their technological applications. Moreover, evaluating the impact of the gate on the microwave losses in integrated devices is crucial for applications in quantum information processing. In this direction, recent studies reported on switching dynamics in Aluminum (Al)-capped Indium Arsenide (InAs) nanowires \cite{Elalaily2021,Elalaily2023,Elalaily2024}, and on time-domain microwave characterization of Niobium (Nb) Dayem bridges \cite{Joint}.

\begin{figure*} [ht!]
        \begin{center}
                \includegraphics [width=1\textwidth]{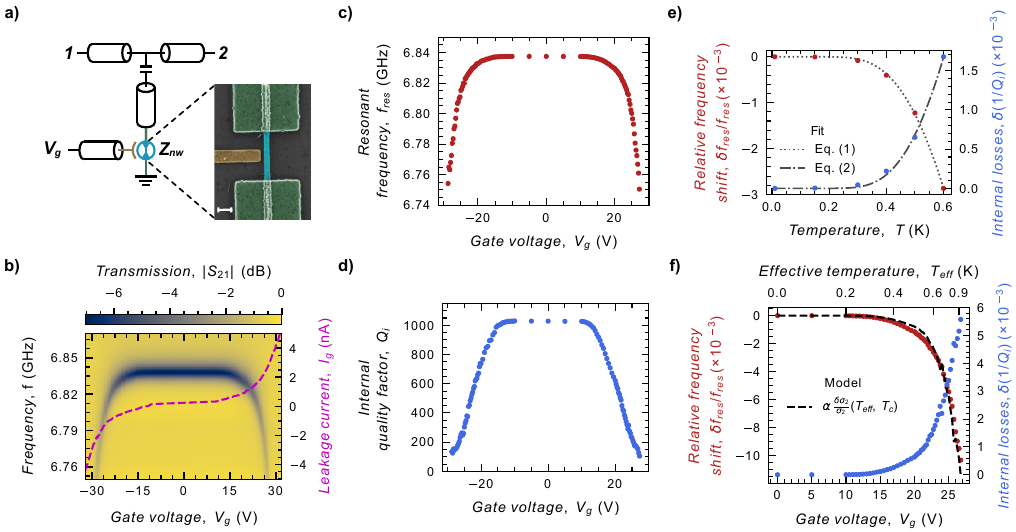}
        \end{center}
        \caption{Device Spectroscopy. \textbf{a)} Simplified schematic of the device. \textit{Zoom-out}: Scanning-electron micrograph of the nanowire. Tha scale bar is 300 nm. \textbf{b)} Heat map of the feedline transmission spectrum, $|S_{21}|$ versus gate voltage, $V_{\rm{g}}$ and frequency, $f$. On the right axis we plot the leakage current $I_{\rm{g}}$ versus $V_{\rm{g}}$. \textbf{c)-d)}  Resonant frequency, $f_{\rm{res}}$ and internal quality factor, $Q_{\rm{i}}$ as a function of $V_{\rm{g}}$.
        \textbf{e)} Temperature-driven data with $V_{\rm{g}}=0$: relative frequency shift $\delta f_{\rm{res}}/f_{\rm{res}}$ (left axis) and inverse quality factor, $\delta(1/Q_{\rm{i}})$ (right axis) are plotted versus temperature. Equations \ref{df} and \ref{dq} are fitted to the data, according to Mattis-Bardeen model. Averaged fit result: $\alpha=(0.17\pm0.01)$ and $T_{\rm{c}}=(1.34\pm0.05)$ K.
        \textbf{f)} Gate-driven data for $T=10$ mK: $\delta f_{\rm{res}}/f_{\rm{res}}$ (left axis) and $\delta(1/Q_{\rm{i}})$ (right axis) versus $V_{\rm{g}}$ (bottom axis). For each data point in $\delta(1/Q_{\rm{i}})$, we obtain an effective temperature $T_{\rm{eff}}$ (top axis), assuming that the gate effect is equivalent to an increase in temperature and using Eq.~\ref{dq} . The predicted $\delta f_{\rm{res}}/f_{\rm{res}}$, based on Eq.~\ref{df} with the parameter obtained form the fit in panel e) and $T=T_{\rm{eff}}$, is plotted as a dashed line).}
        \label{fig1}
\end{figure*}

In this paper, we investigate an InAs nanowire -of the same type discussed in Ref.~\cite{Elalaily2024}- with the microwave techniques used in Ref.~\cite{Joint}. The core of our device is a hybrid semiconducting-superconducting nanowire with impedance $Z_{\rm{NW}}$, consisting of a 150 nm diameter InAs core which is covered with a 20 nm Al full shell [see Fig.~\ref{fig1} a)]. Since the thickness of the Al shell is orders of magnitude larger than the Thomas-Fermi length in Al, the InAs core is screened from the electric field and does not play any role in the gate response of the nanowire \cite{Elalaily2024}, contrary to other configurations \cite{Splitthoff}. The nanowire is embedded in the current antinode of a $\lambda/4$ coplanar waveguide resonator and gated in a finger-type configuration. The resonator is capacitively coupled to the feedline we use for the readout. We perform our measurements in a dilution cryostat with 10 mK base temperature.

We measure the feedline transmission parameter, $S_{21}$, with a vector network analyzer (VNA) as a function of the frequency $f$ and DC gate voltage $V_{\rm{g}}$ applied to the gate. We observe a dip in $|S_{21}|$, corresponding to the resonator mode, which shifts as a function of $V_g$ [see Fig. \ref{fig1} b)]. In particular, the resonator mode shows a sign-symmetric gate response with an onset around $\pm16$ V, together with a rise of gate leakage current up to a few nA. We extract by fit \cite{Probst} [see supplementary material Fig~\ref{supp_spectro}b)-c)] resonant frequency $f_{\rm{res}}$ and internal quality factor $Q_{\rm{i}}$ from the resonator trace for each $V_{\rm{g}}$ [see Fig.~\ref{fig1} c) and d)]. Both $f_{\rm{res}}$ and $Q_{\rm{i}}$ decrease monotonically, from $f_{\rm{res}}=6.84$ GHz and $Q_{\rm{i}}=1000$ at zero gate to $f_{\rm{res}}=6.75$ GHz, $Q_{\rm{i}}=100$ around $V_{\rm{g}}=\pm30$ V.

We interpret our result as consequence of quasiparticles generation in the gated nanowire due to Cooper pair breaking. An increase in the quasiparticle density $n_{\rm{qp}}$, has a double effect on $Z_{\rm{NW}}$: it increases the kinetic inductance $L_{\rm{k}}$, causing a shift in the resonant frequency, $\delta f_{\rm{res}}$, and introduces a parallel dissipative channel, i.e. a conductance $G$, which causes a shift in the internal quality factor, $\delta Q_{\rm{i}}$ \cite{Tinkham}. Note that we assume the nanowire fully contributes to the kinetic component of the device inductance (supplementary material Fig. \ref{supp_spectro}). 

By translating the shift in $Z_{\rm{NW}}$ in terms of a shift in the complex conductivity $\sigma = \sigma_1 + i\sigma_2$, we apply the generalized Mattis-Bardeen model \cite{Mattis, Owen} and rewrite the shift in $f_{\rm{res}}$, $Q_{\rm{i}}$ as

\begin{equation}
    \frac{\delta f_{\rm{res}}}{f_{\rm{res}}} = \frac{\alpha}{2}\frac{\delta\sigma_2}{\sigma_2}
    \label{df}
\end{equation}
and
\begin{equation}
    \delta\left(\frac{1}{Q_{\rm{i}}}\right) = \alpha\frac{\delta\sigma_1}{\sigma_2},
    \label{dq}
\end{equation}

\noindent respectively, where $\alpha$ is the participation ratio between the gate-affected kinetic inductance and the total inductance of the resonator \cite{Ryu, Gao, Hu}. Here, assuming thermal equilibrium, real and imaginary parts $\sigma_{1,2}$ are function of temperature $T$ and critical temperature $T_{\rm{c}}$, i.e. $\sigma_{\rm{i}}=\sigma_{\rm{i}}(T,T_{\rm{c}})$ [supplementary material Eqs. \ref{sigma1}, \ref{sigma2}].

In order to obtain $T_{\rm{c}}$ and $\alpha$, we measure $\delta f_{\rm{res}}$ and $\delta(1/Q_{\rm{i}})$ in a temperature-driven experiment where $V_{\rm{g}}=0$ V, and the increasing in $n_{\rm{qp}}$ is due to the temperature rise in the cryostat. We fit equations \ref{df}, \ref{dq} to our temperature-driven data [see Fig.~\ref{fig1} e)] and obtain $\alpha=0.17$ and $T_{\rm{c}}=1.34$ K, compatible with aluminum thin-film \cite{Chubov,Meservey}.

Then we apply the same model to our gate-driven data, using an effective-temperature approach. After obtaining $T_{\rm{c}}$ and $\alpha$ from the temperature-driven experiment, we use Eq.~\ref{dq} to estimate the effective temperature $T_{\rm{eff}}$ from the measured internal losses $\delta(1/Q_{\rm{i}})$. In turn, by using the obtained $\alpha$, $T_{\rm{c}}$ and $T_{\rm{eff}}$, we find that Eq.~\ref{df} describes the gate-driven shift in frequency [see Fig. \ref{fig1}f)], which validates our analysis. Similar experiments about gate-effect on superconductors in the microwave regime are reported in Refs.~\cite{Catto, Ryu}. However, in Ref.~\cite{Catto}, the Mattis-Bardeen model is used to estimate resonator inductance and resistance from material properties, geometry and temperature, which then feed a simulation that predicts resonant frequency and quality factor. On the other hand, Ref.~\cite{Ryu} applies Eqs.~\ref{df}, \ref{dq} for a remote-gate configuration calculating an effective quasiparticle density and assuming the temperature to be constant. In this case, the model slightly deviates from the experimental data, which the authors attribute to a heating effect. 

To summarize this section, the frequency domain characterization of our device shows a gate-driven frequency tunability in excess of $90$ MHz, with a contextual $\sim85\%$ decrease in internal quality factor, both related to a rise in the leakage current from the gate. Furthermore, we can describe the gate-driven relation between $\delta f_{res}$ and $\delta(1/Q_i)$ with an effective temperature in the framework of the Mattis-Bardeen model.

\begin{figure*} [ht!]
        \begin{center}
                \includegraphics [width=1\textwidth]{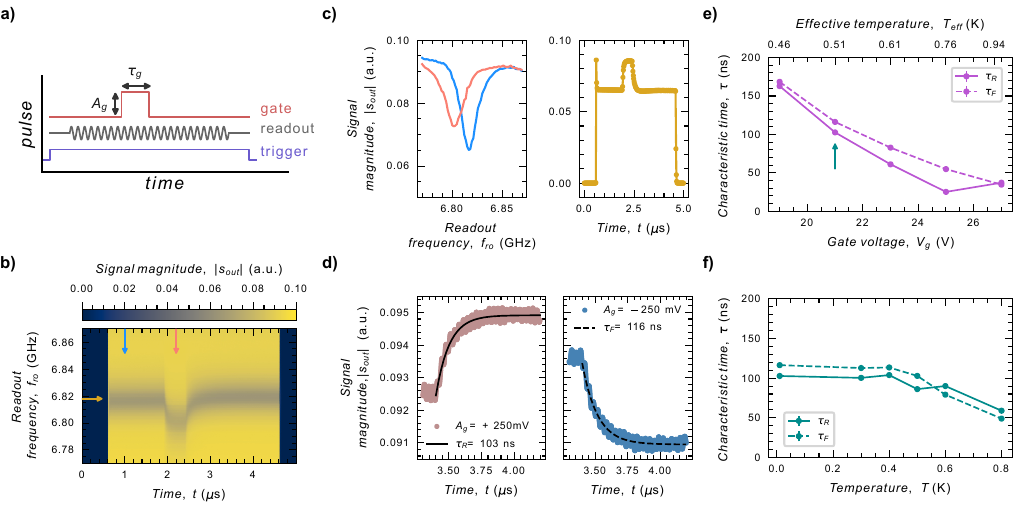}
        \end{center}
        \caption{Device dynamics: time domain spectroscopy. \textbf{a)} Pulse sequence used for the time-domain measurements. \textbf{b)} Heat map of resonator time traces recorded for different readout pulse frequencies $f_{\rm{ro}}$ at gate voltage $V_{\rm{g}}=23.5$ V. With reference to the scheme in panel a), the trigger period is 5 $\mu$s, a 4 $\mu$s long readout pulse starts at 0.6 $\mu$s and a gate pulse pulse with $A_{\rm{g}}=1.5$ V and $\tau_{\rm{g}}=500$ ns starts around 2 $\mu$s. Arrows indicate line cuts discussed in panel c). \textbf{c)} Line cuts from panel b). \textit{Left}: vertical line cuts at 1 $\mu$s (blue) and 2.2 $\mu$s (red) showing the resonator shift due to the gate pulse. \textit{Rigth}: horizontal line cut at 6.815 GHz showing resonator ring-up and gate response in time domain. \textbf{d)} Time traces at $V_{\rm{g}}=21$ V showing resonator rising (left) and falling (right) response under positive and negative gate pulses, respectively. Gate parameters are $|A_{\rm{g}}|=250$ mV and $\tau_{\rm{g}}=800$ ns. Rising (continuous line) and falling (dashed line) exponential function are fitted to the gate response, obtaining characteristic time $\tau_{\rm{R}}$ and $\tau_{\rm{F}}$, respectively.  \textbf{e)} Rising and falling time as a function of $V_{\rm{g}}$ and effective temperature $T_{\rm{eff}}$. The green arrow indicates the gate working point used for the temperature-driven data reported in panel f). \textbf{f)} Rising and falling time as a function of the temperature at a fixed $V_{\rm{g}}=21$ V.}
        \label{fig2}
\end{figure*}

We now investigate the dynamic response of our system. We discuss two different techniques for such a measurement: pulsed and continuous gate-excitation. In both cases,  we send our gate excitation on top of a DC working point $V_{\rm{g}}$, which is above the onset of the gate effect. Excitation and DC signals are combined to the same gate line at the mixing chamber level via a bias-tee [see supplementary material Fig.~\ref{supp_setup}a)].   

Pulsed measurements consist of three pulses: a trigger pulse, a readout pulse, and a gate pulse [see Fig.~\ref{fig2}a)]. We generate all pulses with a microwave digital transceiver, Presto \cite{presto}. The trigger pulse defines the time span during which we record the output signal $s_{\rm{out}}$ from the feedline. While the trigger is on, we send a readout pulse with frequency $f_{\rm{ro}}$ to the feedline input and a square pulse with amplitude $A_{\rm{g}}$ and duration $\tau_{\rm{g}}$ to the gate line. We average $s_{\rm{out}}$ over $10^5-10^6$ repetition of the triggered sequence and repeat the measurement for different $f_{\rm{ro}}$, obtaining a 2D map $s_{\rm{out}}(t, f_{\rm{ro}})$ [see Fig.~\ref{fig2}b)]. 

When $f_{\rm{ro}}\sim f_{\rm{res}}$, the readout signal excites the resonator, leaving a signature of the resonant frequency and bandwidth in $s_{\rm{out}}$. Consequently, the data at a given time, $s_{\rm{out}}(t=t_{\rm{x}}, f_{\rm{ro}})$, provides an ``snapshot'' of the resonator's spectroscopy at time $t_x$ [see left side of Fig.~\ref{fig2}c)]. In particular, we fit Lorentzian functions to these frequency traces [supplementary material Fig.~\ref{supp_td}a)], from which we obtain the instantaneous $f_{\rm{res}}$ and loaded quality factor $Q_{\rm{L}}$. By comparing $\delta f_{\rm{res}}$ due to the gate pulse at steady-state with the VNA measurements discussed in the previous section [see left side of Fig.~\ref{fig2}c) versus Fig. \ref{fig1}c)], we calibrate the attenuation of the RF line we use to drive the gate, obtaining the effective pulse amplitude. 

On the other hand, individual time traces $s_{\rm{out}}(t, f_{\rm{ro}}\sim f_{\rm{res}})$ describe the dynamics of our system, namely the resonator ring-up spike around $t=0.6\;\mu$s and the gate response around $t=2\;\mu$s [see right side of Fig.~\ref{fig2}c)]. The ring-up decays exponentially to a steady-state level with a characteristic time that defines the resonator response time \cite{Heidler}. From an exponential fit [see supplementary material Fig.~\ref{supp_td}b)], we obtain $\tau_{\rm{res}}=10$ ns, which is compatible with $Q_{\rm{L}}/(2\pi f_{\rm{res}})=16$ ns obtained from the Lorentzian fits.

We then fix $f_{\rm{ro}}\sim f_{\rm{res}}$ and choose $A_{\rm{g}}=250$ mV, which induces a variation in $|s_{out}|$ up to a few percents while still leaving the system in the linear response regime (see supplementary material). Further, we set 500 ns$\;\le\tau_{\rm{g}}\le800$ ns, which is long enough to allow $|s_{out}|$ to saturate to a steady value during the pulse excitation but short enough to avoid distortions due to the bias-tee (see supplementary material). With these pulse settings, we investigate the characteristic time of the gate response in the resonator time trace as a function of $V_{\rm{g}}$ and cryostat base temperature $T$. In particular, for each $V_{\rm{g}}$ or $T$, we measure the response for both positive and negative pulse ($\pm A_{\rm{g}}$). We fit exponential functions to the rising and falling edges of the positive and negative pulse-response shape, respectively, obtaining rising and falling characteristic times, $\tau_{\rm{R,F}}$ [see Fig. \ref{fig2}d)]. We discard the falling (rising) part of the positive (negative) pulse response which are affected by the bias-tee in the gate line [supplementary material Fig. \ref{supp_td}d)]. Here, rising (falling) refers to a positive (negative) change in the gate voltage, respectively, which does not necessarily corresponds to an increase (decrease) in $|s_{\rm{out}}|$.

Since we work in linear regime, we find no clear difference between $\tau_{\rm{R}}$ and $\tau_{\rm{F}}$: they both decrease monotonically with increasing $V_{\rm{g}}$ [see Fig. \ref{fig2}e)], ranging from 170 ns at $V_{\rm{g}}=19$ V to 35 ns at $V_{\rm{g}}=27$ V. We then fix $V_{\rm{g}}=21$ V and measure the characteristic times by stepping the temperature up to 800 mK. We do not observe any clear different behavior between $\tau_{\rm{R}}$ and $\tau_{\rm{F}}$ in this case, either: they remain constant at around 130 ns up to 500 mK, then monotonically decrease to 50 ns at 800 mK [see Fig. \ref{fig2}f)]. Even in time domain, temperature-driven results are compatible with the effective temperature approach. In fact, as discussed in the previous section, $V_{\rm{g}}=21$ V corresponds to $T_{\rm{eff}}= 510$ mK, which is compatible with the onset in the temperature-induced decrease in $\tau_{\rm{R,F}}$ for $T\gtrsim500$ mK. 

We compare our data with the expected quasiparticle recombination time, $\tau_{\rm{qp}}\sim\tau_{\rm{0}}/n_{\rm{qp}}$, where $\tau_{\rm{0}}$ is the electron-phonon interaction time in Al \cite{Visser}. By using the thermal equilibrium $n_{\rm{qp}}(T_{\rm{c}},T)$, with the critical temperature, $T_{\rm{c}}$ obtained from the temperature-driven frequency spectroscopy, we find that $\tau_{\rm{0}}=30$ ns leads to a $\tau_{\rm{qp}}$ ranging from 100 ns at $T=500$ mK to 10 ns at $T=1$ K (see supplementary material), which is compatible with the characteristic times we obtain in our experiment as a function of $T_{\rm{eff}}$. Whereas $\tau_{\rm{0}}$ is often measured to be on the order of 450 ns in aluminum \cite{Visser,Kaplan, Martinis}, 30 ns is still within the reported range of values for high concentration of aluminum oxide impurities film \cite{Chi}. On the other hand, also an increase in phonon or quasiparticle density due to nonequilibrium process can effectively decrease $\tau_{\rm{0}}$ and $\tau_{\rm{qp}}$ \cite{Chi, Visser, Goldie}.


\begin{figure} [!htbp]
        \includegraphics[width=\columnwidth]{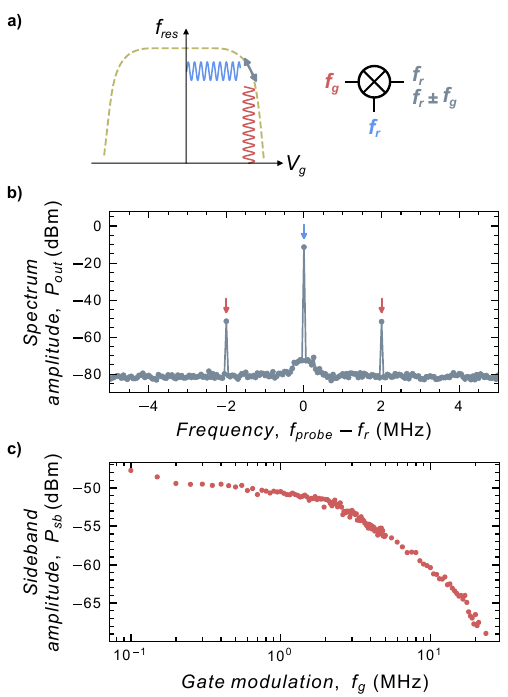}
        \caption{Device dynamics: sidebands spectroscopy. \textbf{a)} Schematic of the measurement. \textit{Left:} Continuous gate excitation modulates the resonator response due to the gate-dependence of the resonator frequency, $f_{\rm{res}}(V_{\rm{g}})$. \textit{Right:} Mixer-equivalent circuit of our system under continuous modulation. \textbf{b)} Sideband excitation for $V_{\rm{g}}=27$ V, $f_{\rm{g}}=2$ MHz and $P_{\rm{g}}=-5$ dBm. Resonator and sidebands peaks are indicated with blue and red arrows, respectively. \textbf{c)} Averaged upper and lower sidebands amplitude [see red arrows in panel b)] as a function of gate modulation.}
        \label{fig3}
\end{figure}


We now perform a continuous-gate excitation around $V_{\rm{g}}=27$ V by swapping our pulses with continuous tones. In particular, we send a resonant tone $f_{\rm{r}}$ to the feedline input and a tone $f_{\rm{g}}$ to the gate, which modulates the device impedance and, consequently, the resonator response. This configuration is equivalent to a mixer where $f_{\rm{r}}$ and $f_{\rm{g}}$ play the role of local oscillator and intermediate frequency, respectively [see Fig. \ref{fig3}a)].

We observe sidebands at $f_{\rm{r}}\pm f_{\rm{g}}$ in the feedline output signal for $f_{\rm{g}} = 2$ MHz [see Fig. \ref{fig3}b)], which proves that $f_{\rm{g}}$ modulates $f_{\rm{r}}$. The amplitude of the sideband peaks decreases monotonically with increasing $f_{\rm{g}}$ [see Fig. \ref{fig3}c)]: it ranges from -50 dBm at low frequency to -68 dBm at $f_{\rm{g,max}}=25$ MHz with a -3 dB point $f_{\rm{g,-3dB}}=3$ MHz, from which we estimate an equivalent characteristic response time of $(2\pi f_{\rm{g,-3dB}})^{-1}=50$ ns \cite{Oppenheim}, on the same order of what obtained in the time domain. Sidebands result, which is not limited by bias-tee (bias-tee cutoff is $40\;$kHz$\;< f_{\rm{g}}$), validates time domain experiments. 

In summary, we discuss the gate response of a Al-capped InAs nanowire embedded in a quarter-wavelength microwave resonator. Spectroscopy characterization of such a system shows a frequency tunability of the order of 90 MHz at the price of a degradation in $Q_{\rm{i}}$ from 1000 to 100. The onset of the gate response  coincides with a rise of the leakage current up to a few nA. We describe the system by the Mattis-Bardeen model with an effective temperature approach, suggesting that the gate response in our device is explained by a heating effect. We note, however, that our use of an effective temperature to model the change in the nanowire impedance does not imply  that the quasiparticle population is in thermal equilibrium. In fact, other studies suggest that gating effects on superconductors are related to out-of-equilibrium processes \cite{Pashkin, RitterPhonon, Elalaily2023, Elalaily2024}. We study the dynamic response of the resonator under both pulsed and continuous gate-excitation. Pulsed excitation is performed in time domain and it shows a gate-dependent response time down to 35 ns. Even in this experiment, a comparison with temperature-driven measurements suggests that the effect is related to a temperature increase. Continuous excitation confirms the values obtained in time-domain, showing sidebands with a -3 dB point at 3 MHz, which gives a characteristic modulation time of 50 ns. Comparison with the expected quasiparticle recombination time, suggests that our time domain results agree with the quasiparticle dynamics if we take into account oxygen contamination in our aluminum film or nonequilibrium effects.

Our findings, together with Ref.~\cite{Joint}, demonstrate gate-response performances down to tens of nanoseconds, which makes gate-controlled superconducting devices a promising candidate for superconducting switch and cryoelectronics \cite{Ruf_rev}. Moreover, further investigation of the gate dynamics in materials with lower electron-phonon characteristic time and higher critical temperature, such as Tantalum (Ta), Titanium-Nitride (TiN) and Niobium-Titaniu-Nitride (NbTiN) \cite{ Visser, Hu, Wilson, Rooij}, could potentially improve the response time. 

\section*{Acknowledgments}

The discussed device was partially fabricated in the Chalmers Myfab cleanroom facility. This work was financially supported by the European Research Council via Grant No. 964398 SuperGate, OTKA
K138433 and EIC Pathfinder Challenge QuKiT (101115315). This paper was supported by the János Bolyai Research Scholarship of the Hungarian Academy of Sciences and by the EK\"OP-24-4-II-BME-95 University Research Scholarship Program of the Ministry for Culture and Innovation from the source of the National Research, Development and Innovation Fun. This research was also supported by the Ministry of Culture and Innovation and the National Research, Development and Innovation Office within the Quantum Information National Laboratory of Hungary (Grant No. 2022-2.1.1-NL-2022-00004), and Novo Nordisk Foundation SolidQ. This work is supported by the Knut and Alice Wallenberg foundation via the Wallenberg Centre for Quantum Technology (WACQT). SG acknowledges financial support from the European Research Council via Grant No.~101041744 ESQuAT.


\newpage

\onecolumngrid

\renewcommand{\thefigure}{S\arabic{figure}}
\setcounter{figure}{0} 
\renewcommand{\theequation}{S\arabic{equation}}
\setcounter{equation}{0} 

\section*{Supplementary Material}

\subsection{Device}

\begin{figure*} [ht!]
        \begin{center}
                \includegraphics [width=0.9\textwidth]{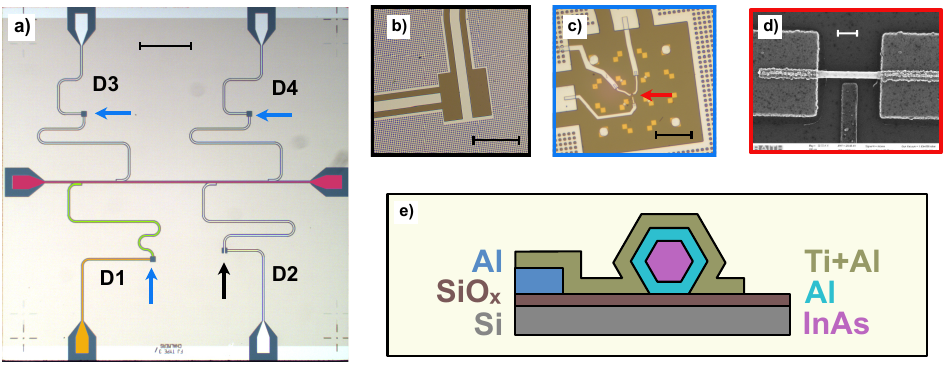}
        \end{center}
        \caption{Device fabrication. \textbf{a)} False-colored optical picture of a chip lithographically identical to the one that was used in the experiments. The feedline (red) is capacitively coupled to four devices (D1, D2, D3 and D4) consisting of a $\lambda/4$ CPW resonator and a gate line each. The resonator and gate line are colored in green and yellow, respectively, for the device discussed in the main text, D1. Blue arrows indicate the gating area where NWs are embedded in the resonator line. A black arrow indicates the reference resonator without NW. The scale bar is 1 mm. \textbf{b)} Optical zoom-in on the bare resonator. The scale bar is 500 $\mu$m. \textbf{c)} Optical zoom-in on gating area with showing a NW (red arrow) shunting the resonator line to the ground plane and the gate line tapered towards the NW. The scale bar is 30 $\mu$m. \textbf{d)} SEM picture of the NW [same as Fig. \ref{fig1} a) in main text]. The scale bar is 400 nm. \textbf{e)} Material stack used in the fabrication of the sample. }
        \label{supp_fab}
\end{figure*}

The device discussed in the main text is part of a 7x7 mm$^2$ chip [see Fig.~\ref{supp_fab}a)]. The chip hosts four resonator-gate devices and we name them D1, D2, D3 and D4. Device D2 is a reference resonator without nanowire (NW) [see Fig.~\ref{supp_fab}b)], while for D1, D3 and D4 the resonator is shunted to ground via a NW [see Fig.~\ref{supp_fab}c) and d)]. 
The sample is fabricated on a Si substrate [see Fig.~\ref{supp_fab}e)]. First, we thermally grow a 200 nm-thick oxide layer on a 2-inch high-resistivity silicon wafer using a furnace. Next, we evaporate a 150 nm-thick aluminum film across the entire wafer. Using a maskless UV lithography system, we pattern the microwave circuits, including feedlines, gate lines, and resonators. The patterns are transferred to the aluminum layer by immersing the wafer in an aluminum etchant solution for approximately two minutes. Subsequently, we pattern alignment markers using e-beam lithography, followed by the lift-off of a 25 nm-thick gold layer. This step prepares the wafer for the deposition and integration of nanowires with the existing circuitry. We deposit the NWs on the sample with a nano-manipulator and then embedded in the cQED structure with 100 nm of Ti-Al  contacts. We obtain the contact via lif-off process and patterned with EBL. Before evaporation, we clean the pattern with Ar milling to reduce contact resistance. 

\subsection{Setup}
After fabrication, we wire-bond our sample to a PCB with Al wires and close it inside a light-tight RF box. The box is thermally anchored to the mixing chamber plate of a dilution cryostat with base temperature of 10 mK [see Fig. \ref{supp_setup}a)]. Gate lines, DC and RF, are combined with a bias-tee with 40 kHz cutoff. Both the input line and the gate lines are attenuated and filtered through the cryostat stages; the output signal is amplified both at 4K and at room temperature. We characterize the sample in multiple cooldowns. Initially, we connect all the gate lines to the same bias-tee via a switch installed at the mixing chanber lavel. In later cooldowns, we terminate all the gate lines but D1, which is the only one showing showing gate response (see Sec.~\ref{supp_spectr}). Regarding the measurements:
\begin{itemize}
    \item DC gate voltage $V_{\rm{g}}$ is controlled by a Stanford Research DC205 voltage source and the leakage current is obtained from the voltage drop across a 4.6 k$\Omega$ series resistance, measured with a Keithley 2000 Multimeter [see Fig. \ref{supp_setup}b)].
    \item Spectroscopy measurement  are performed with port 1 and port 2 of a Keysight P5024A VNA with the gate RF line teminated with 50 $\Omega$ [see Fig. \ref{supp_setup}c)].
    \item Time domain measurements with squared pulse gate-excitation are performed with a digital transceiver with 1 GS/s sampling rate, Presto \cite{s_presto}, which generates both readout and gate pulses, and analyses the output signal from the feedline [see Fig. \ref{supp_setup}d)]. In particular, input (output) readout signals are generated (recorded) by digital up- (down-) conversion with 100-125 MHz intermediate frequency. Down-converted output signal is then demodulated and filtered, to eliminate spurious modes. The resulting time limitation from the instrument is 8 ns. Room temperature gate-pulse amplifier has 2 ns response-time. 
    \item Sidebands are measured with Keysight P5024A. On the one hand, port 1 and port 3 are in continuous wave (CW) mode and generate readout tone and gate excitation, respectively. On the other hand, port 2 is in spectrum analyzer mode (SA) and analyses the output signal. 
    \end{itemize}    

\subsection{Spectroscopy measurements}
\label{supp_spectr}

\begin{figure*} [ht!]
        \begin{center}
                \includegraphics [width=1\textwidth]{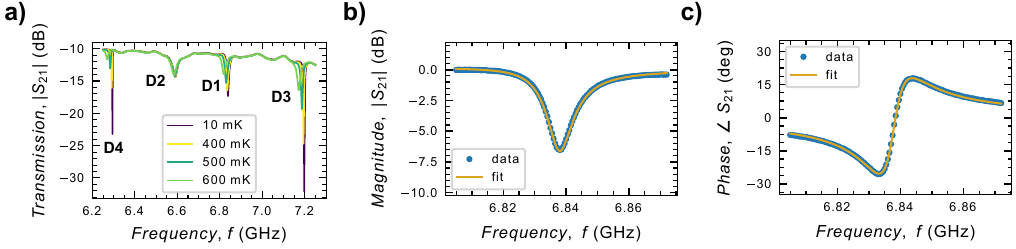}
        \end{center}
        \caption{Frequency spectroscopy. \textbf{a)}: Wide frequency-span feedline transmission spectra $S_{21}$ for different temperatures at $V_{\rm{g}} =0$ V. \textbf{b)}-\textbf{c)} We show D1 resonator amplitude and phase at 10 mK for $V_{\rm{g}}=0$ V. We perform the fit following Ref.~\cite{s_Probst}, obtaining $f_{\rm{res}}=6.837$ GHz, $Q_{\rm{i}}=980$ and $Q_{\rm{c}}=828$.}
        \label{supp_spectro}
\end{figure*}

A broad scan of the transmission $S_{21}$ versus frequency at different temperatures shows three modes shifting towards lower frequency for $T>300$ mK (6.3 GHz, 6.8 GHz and 7.2 GHz) and a stable mode at 6.6 GHz [see Fig. \ref{supp_spectro}a)]. The latter corresponds to D2, the bare resonator, while the formers correspond to the NW-loaded ones, D1, D3 and D4. Since the frequency shift as a function of temperature is due to the kinetic inductance \cite{s_Tinkham}, we can conclude that in our devices the kinetic contribution of the inductance is in the NWs, while the resonator lines give geometric contribution. This is expected also from the different thicknesses between CPW Al layer (150 nm) and NW Al layer (20 nm) \cite{s_Tinkham}. The measured resonant frequencies for D1, D3 and D4, are compatible with Sonnet simulations of our resonators in which we assume $\sim250$ pH contribution form the NWs, on the same order of what reported in \cite{s_Splitthoff} for similar NWs. We obtain resonant frequency and quality factors for the D1 resonator by using the ``circle-fit'' method \cite{s_Probst} [see Fig~\ref{supp_spectro}b)-c)].

\subsection{Mattis-Bardeen model}

Regarding our analysis of $\delta f_{\rm{res}}$ and $\delta(1/Q_{\rm{i}})$ within the Mattis-Bardeen model, equation \ref{df} and \ref{dq} in the main text require an expression for $\sigma_{\rm{i}}$. At low temperature $T$ and small frequency $\omega$, i.e. $k_B T<<\Delta$ and $\hbar\omega<<\Delta$, with $\Delta$ being the superconducting gap and $k_B$, $\hbar$ being the Boltzmann and reduced Plank constants, respectively, $\sigma_{1,2}$ can be written as \cite{s_Ryu}
\begin{equation}
    \frac{\sigma_1}{\sigma_n}=\frac{2\Delta_{\rm{0}}}{\hbar \omega}\frac{n_{\rm{qp}}}{N_{\rm{0}}\sqrt{2\pi k_B T\Delta_{\rm{0}}}}\sinh{\left(\xi\right)}K_{\rm{0}}(\xi),
    \label{sigma1}
\end{equation}
and
\begin{equation}
    \frac{\sigma_2}{\sigma_n}=\frac{\pi\Delta}{\hbar\omega}\left[1-\frac{n_{\rm{qp}}}{2N_{\rm{0}}\Delta_{\rm{0}}}\left(1+\sqrt{\frac{2\Delta_{\rm{0}}}{\pi k_BT}}e^{-\xi}I_{\rm{0}}(\xi)\right)\right],
    \label{sigma2}
\end{equation}
where $\sigma_n$ is the normal state conductivity, $N_{\rm{0}}$ is the electrons density of states at the Fermi level, $\Delta_{\rm{0}}=1.764k_BT_{\rm{c}}$ is the superconducting gap at $T=0$ K, $\xi=\hbar\omega/2k_BT$ and $K_{\rm{0}}$, $I_{\rm{0}}$ are the Bessel function of first and second order, respectively.

Equations \ref{sigma1}, \ref{sigma2} take into account an increase in the quasi-particles density due to generic pair-breaking. In case of thermal excitation we can substitute $n_{\rm{qp}}$ with

\begin{equation}
    n_{\rm{qp}} = 2N_{\rm{0}}\sqrt{2\pi k_BT\Delta_{\rm{0}}}\;\exp{\left({-\frac{\Delta_{\rm{0}}}{k_BT}}\right)}.
    \label{qpT}
\end{equation}

\subsection{Time-domain measurements}

\begin{figure*} [ht!]
        \begin{center}
                \includegraphics [width=1\textwidth]{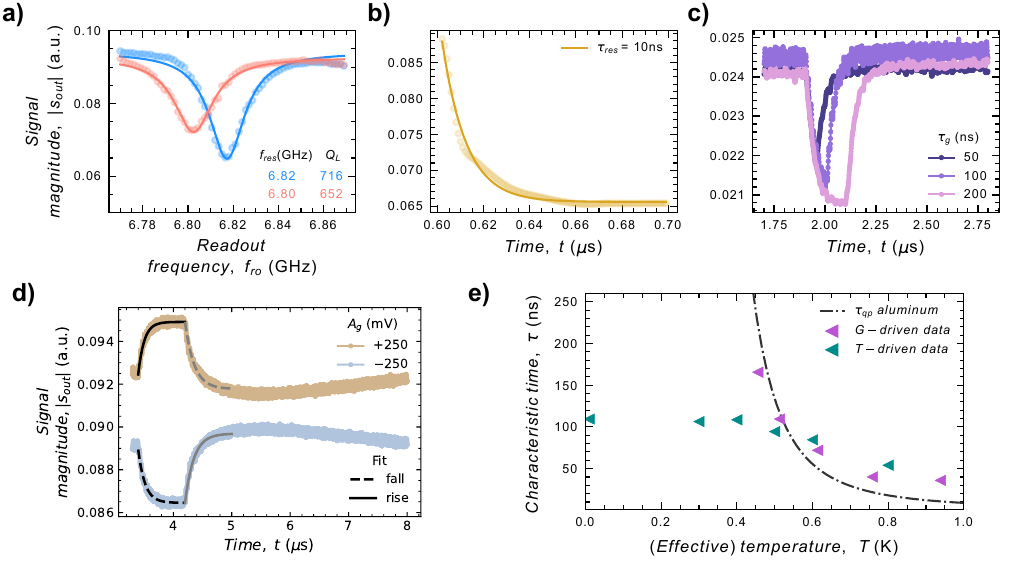}
        \end{center}
        \caption{Time domain: preliminary measurements and comparison with quasiparticle recombination time. \textbf{a)} Lorentzian fits [Eq. \ref{lorentzian}] of resonant frequencies obtained with time domain spectroscopy [see left-side of Fig.~\ref{fig2}c) in the main text]. \textbf{b)} Exponential fit [Eq. \ref{decay}] of resonator ring-up [see right-side of Fig. \ref{fig2}c) in the main text]. \textbf{c)} Comparison of gate response for different gate pulse duration $\tau_{\rm{g}}$. \textbf{d)} Comparison of gate response positive ($+A_{\rm{g}}$) and negative ($-A_{\rm{g}}$) amplitude showing the long-time response of the resonator after the gate pulse. Curves are vertically shifted for clarity. Exponential fit of the rising and falling parts of the gate response are shown in continuous and dashed lines, respectively. Considered and discarded fit results are shown in black and gray, respectively. In this measurement we are at base temperature, 10 mK, and $V_{\rm{g}}=21$ V.  \textbf{e)} Comparison with the quasiparticle recombination time: orchid and cyan data points represent average characteristic times $(\tau_R+\tau_F)/2$ obtained at 10 mK as a function of DC working point, $V_g$, and at $V_g=21$ V as a function of temperature, $T$, respectively. The voltage has been converted to effective temperature for the gate-driven data, as explained in the main text. The gray dash-dot line represents the expected quasiparticle recombination time obtained from Eq.~\ref{tau_qp_T} with $\tau_0=30$ ns.}
        \label{supp_td}
\end{figure*}

We fit Lorentzian functions to the resonant curves obtained from time domain measurements [see Fig.~\ref{supp_td}a)]:
\begin{equation}
    s_{\rm{out}}(f)= \beta\frac{\gamma^2}{(f-f^*_{\rm{res}})^2+\gamma^2}+\theta,
    \label{lorentzian}
\end{equation}
where $f_{\rm{res}}$ is the resonant frequency, $\gamma$ is the FWHM of the curve, i.e. $Q_{\rm{L}}=f_{\rm{res}}/\gamma$, $\beta$ is the amplitude of the dip and $\theta$ is the offset of the baseline. We fit exponential decay functions to the resonator ring-up [see Fig.~\ref{supp_td} b)] and rising/falling parts of the gate response [see Fig.~\ref{supp_td} d)]:
\begin{equation}
    s_{\rm{out}}(t)= A+(B-A)e^{(t-t_{\rm{0}})/\tau},
    \label{decay}
\end{equation}
where $A, B$ are the initial, final value of the exponential, respectively, $t_{\rm{0}}$ if the offset on the starting time and $\tau$ is the characteristic decay time. $B$ and $\tau$ are used as fit parameters.

For our test pulse, we choose the gate duration such that $200\;\rm{ns}<\tau_{\rm{g}}\le1\;\rm{\mu s}$, which ensures a pulse long enough to saturate the gate response [see Fig.~\ref{supp_td}c)] and short enough not to be affected by the bias-tee. The falling (rising) part of the positive (negative) pulse triggers a slow response of the output signal [see Fig.~\ref{supp_td}d)] that we attribute to the frequency-dependent response of the gate line. To avoid the slow-response influence, we discuss results from the rising and falling edges of the positive and negative pulse, respectively, disregarding the others two edges. Note that, as discussed in the main text, $A_{\rm{g}} =\pm250\;\rm{mV}$ gives a symmetric gate response, $\delta s_{\rm{out}}(+A_{\rm{g}})\sim -\delta s_{\rm{out}}(-A_{\rm{g}})$ [see Fig. \ref{supp_td}d)], confirming that the perturbation induced by the gate pulse is in the linear response regime.

The expected quasiparticle recombination time can be written as \cite{s_Visser}:
\begin{equation}
    \tau_{\rm{qp}} = \frac{\tau_{\rm{0}}}{n_{\rm{qp}}}\frac{N_0(k_BT_{\rm{c}})^3}{2\Delta_0^2},
    \label{tau_qp}
\end{equation}
where $\tau_{\rm{0}}$ is the material-dependent electron-phonon interaction time. In case of thermal equilibrium, we can substitute $n_{\rm{qp}}$ with Eq. \ref{qpT} and obtain:

\begin{equation}
    \tau_{\rm{qp}} = \frac{\tau_{\rm{0}}}{\sqrt{\pi}}\left(\frac{k_BT_{\rm{c}}}{2\Delta}\right)^{5/2}\sqrt{\frac{T_{\rm{c}}}{T}}e^{\Delta/k_BT}.
    \label{tau_qp_T}
\end{equation}
\noindent We find that Eq.~\ref{tau_qp_T} with $\tau_0=30$ ns is in agreement with measured characteristic times [see Fig.~\ref{supp_td}e)].

\begin{figure*} [ht!]
        \begin{center}
                \includegraphics [width=0.9\textwidth]{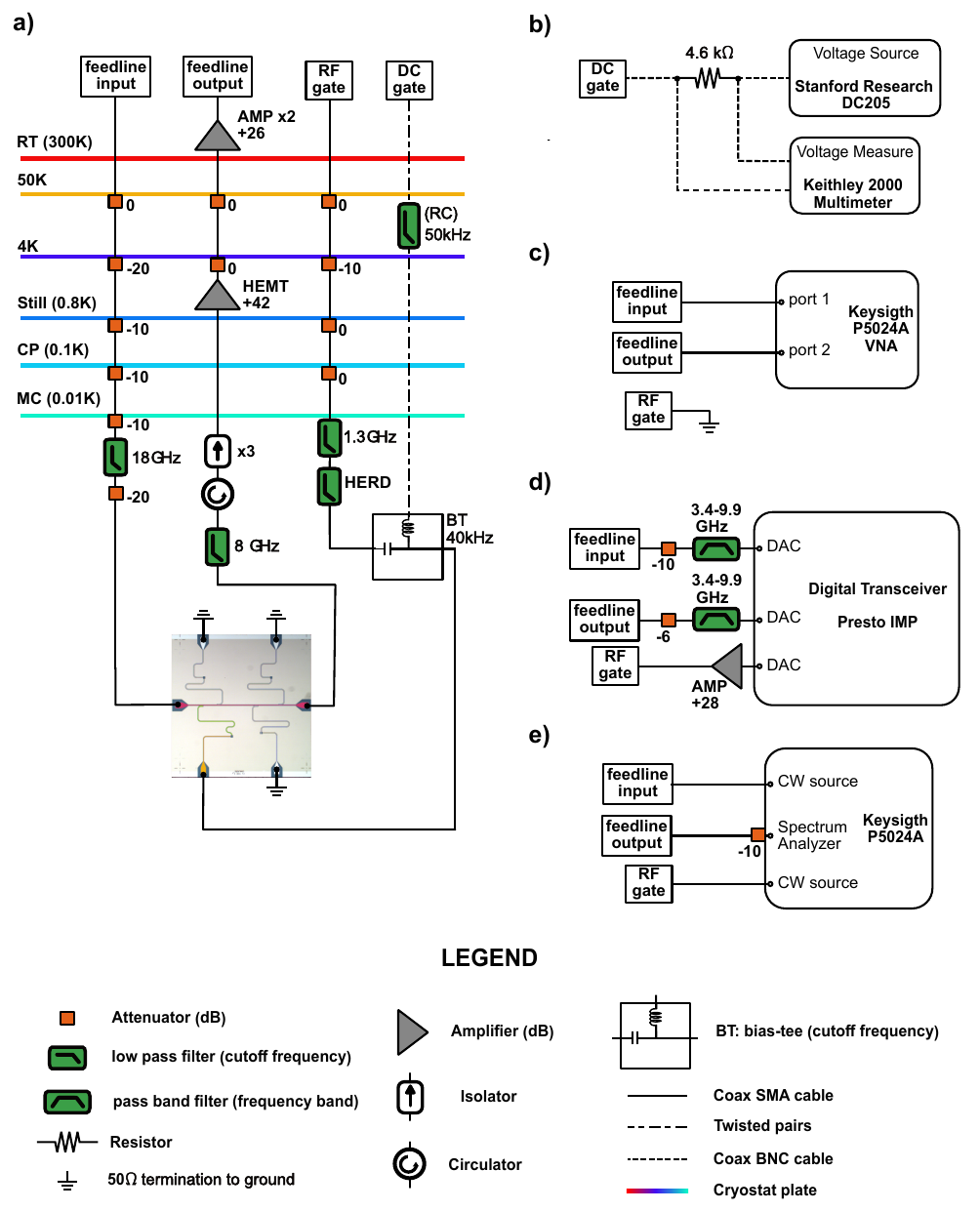}
        \end{center}
        \caption{Experimental setup. \textbf{a)} Simplified schematic of the of the cryostat where we perform the measurements. \textbf{b)} Scheme of the DC gate source and leakage current measurement. \textbf{c)} Scheme of the spectroscopy experiment. \textbf{d)} Scheme of the the time domain experiment. \textbf{e)} Scheme of the sideband experiments.}
        \label{supp_setup}
\end{figure*}


\begin{thebibliography}{10}

\bibitem{Braginski}
A. I. Braginski, "Superconductor Electronics: Status and Outlook", \href{https://link.springer.com/article/10.1007/s10948-018-4884-4}{J Supercond Nov Magn 32, 23–44 (2019))}  .

\bibitem{Kjaergaard}
M. Kjaergaard, M. E. Schwartz, J. Braumüller et al., "Superconducting Qubits: Current State of Play", \href{https://www.annualreviews.org/content/journals/10.1146/annurev-conmatphys-031119-050605?utm_source=chatgpt.com#}{Annu. Rev. Condens. Matter Phys. 11:369–95 (2020)}.  

\bibitem{Tinkham}
M. Tinkham, "Introduction to Superconductivity" (2nd ed.), Dover Publication.

\bibitem{Devoret}
J. Koch, T.M. Yu, J. Gambetta et al., \textit{Charge-insensitive qubit design derived from the Cooper pair box}, \href{https://journals.aps.org/pra/abstract/10.1103/PhysRevA.76.042319}{Phys. Rev. A 76, 042319 (2007)}

\bibitem{Fagaly}
R. L. Fagaly, \textit{Superconducting quantum interference device instruments and applications}, \href{https://pubs.aip.org/aip/rsi/article/77/10/101101/1017955}{Rev. Sci. Instrum. 77, 101101 (2006)}.


\bibitem{Li}
J. Li, P. Barry, T. Cecil et al., \textit{Flux-cooupled tunable superconducting resonator}, \href{https://journals.aps.org/prapplied/abstract/10.1103/PhysRevApplied.22.014080}{Phys. Rev. Applied 22, 014080 (2024)}.

\bibitem{Chen}
Y. Chen, D. van Driel, C. Lampadaris et al., \textit{Gate-tunable superconductivity in hybrid InSb–Pb nanowires}, \href{https://pubs.aip.org/aip/apl/article/123/8/082601/2907410}{Appl. Phys. Lett. 123, 082601 (2023)}.

\bibitem{Burkard}
G. Burkard, M. J. Gullans, X. Mi and J. R. Petta, \textit{Superconductor–semiconductor hybrid-circuit quantum electrodynamics}, \href{https://www.nature.com/articles/s42254-019-0135-2}{ Nat Rev Phys 2, 129–140 (2020)}.

\bibitem{DeSimoni}
G. De Simoni, F. Paolucci, P. Solinas et al., \textit{Metallic supercurrent field-effect transistor}, \href{https://www.nature.com/articles/s41565-018-0190-3}{ Nature Nanotechnology volume 13, 802–805 (2018)}.

\bibitem{Paolucci2018}
F. Paolucci, G. De Simoni, E. Strambini et al., \textit{Ultra-Efficient Superconducting Dayem Bridge Field-Effect Transistor},  \href{https://pubs.acs.org/doi/10.1021/acs.nanolett.8b01010}{ Nano Lett. 18, 7, 4195–4199 (2018)}.

\bibitem{Paolucci2019}
F. Paolucci, F. Vischi, G. De Simoni et al., \textit{Field-Effect Controllable Metallic Josephson Interferometer},  \href{https://pubs.acs.org/doi/full/10.1021/acs.nanolett.9b02369}{ Nano Lett., 19, 9, 6263–6269 (2019)}.

\bibitem{Puglia}
C. Puglia G. De Simoni and F. Giazotto, \textit{Electrostatic Control of Phase Slips in Ti Josephson Nanotransistors},  \href{https://journals.aps.org/prapplied/abstract/10.1103/PhysRevApplied.13.054026}{ Phys. Rev. Applied 13, 054026 (2020)}.

\bibitem{Yu}
S. Yu, L. Chen, Y. Pan, et al., \textit{Gate-Tunable Critical Current of the Three-Dimensional Niobium Nanobridge Josephson Junction},  \href{https://pubs.acs.org/doi/10.1021/acs.nanolett.3c02015}{Nano Lett. 23, 17, 8043–8049 (2023)}.

\bibitem{Pashkin}
I. Golokolenov, A.Guthrie, S. Kafanov et al., \textit{On the origin of the controversial electrostatic field effect in superconductors}, \href{https://www.nature.com/articles/s41467-021-22998-0}{Nat Commun 12, 2747 (2021)}.

\bibitem{RitterPhonon}
M.F. Ritter, N. Crescini, D.Z. Haxell et al, \textit{Out-of-equilibrium phonons in gated superconducting switches},  Nat. Electron., 2022. Available:  \href{https://www.nature.com/articles/s41928-022-00721-1}{Nat Electron 5, 71–77 (2022}.

\bibitem{Catto}
G. Catto, W. Liu, S. Kundu et al., \textit{Microwave response of a metallic superconductor subject to a high-voltage gate electrode}, \href{https://www.nature.com/articles/s41598-022-10833-5}{ Sci Rep 12, 6822 (2022)}.

\bibitem{Elalaily2021}
T. Elalaily, O. Kürtössy, Z. Scherübl, et al., \textit{Gate-Controlled Supercurrent in Epitaxial Al/InAs Nanowires},  \href{https://pubs.acs.org/doi/full/10.1021/acs.nanolett.1c03493}{Nano Lett. 21, 22, 9684–9690 (2021)}.

\bibitem{Elalaily2023}
T. Elalaily, M. Berke, M. Kedves et al., \textit{Signatures of Gate-Driven Out-of-Equilibrium Superconductivity in Ta/InAs Nanowires},  \href{https://pubs.acs.org/doi/10.1021/acsnano.2c10877}{ACS Nano 17, 6, 5528–5535 (2023)}.

\bibitem{Elalaily2024}
T. Elalaily, M. Berke, I. Lija et al., \textit{Switching dynamics in Al/InAs nanowire-based gate-controlled superconducting switch}, \href{https://www.nature.com/articles/s41467-024-53224-2}{ Nat Commun 15, 9157 (2024)}.

\bibitem{Ruf}
L. Ruf, T. Elalaily, C. Puglia, et al., \textit{Effects of fabrication routes and material parameters on the control of superconducting currents by gate voltage}, \href{https://pubs.aip.org/aip/apm/article/11/9/091113/2911843/Effects-of-fabrication-routes-and-material}{APL Mater. 11, 091113 (2023)}.

\bibitem{RitterSwitch}
M.F. Ritter, A. Fuhrer, D.Z. Haxell et al., \textit{A superconducting switch actuated by injection of high-energy electrons}, \href{https://www.nature.com/articles/s41467-021-21231-2}{ Nat Commun 12, 1266 (2021)}.

\bibitem{Ryu}
Y. Ryu, J. Jeong, J. Suh et al, \textit{Utilizing Gate-Controlled Supercurrent for All-Metallic Tunable Superconducting Microwave Resonators}, \href{https://pubs.acs.org/doi/10.1021/acs.nanolett.3c04080}{Nano Lett. 24, 4, 1223–1230 (2024)}.

\bibitem{Joint}
F. Joint, K.R. Amin, I.P.C. Cools and S. Gasparinetti, \textit{Dynamics of gate-controlled superconducting Dayem bridges}, \href{https://pubs.aip.org/aip/apl/article/125/9/092602/3309663}{Appl. Phys. Lett. 125, 092602 (2024)}.

\bibitem{Ruf_rev}
L. Ruf, C. Puglia, T. Elalaily et al., \textit{Gate control of superconducting current: Mechanisms, parameters, and technological potential}, \href{https://pubs.aip.org/aip/apr/article/11/4/041314/3318440/Gate-control-of-superconducting-current-Mechanisms}{App. Phys. Rev. 11, 041314 (2024)}.

\bibitem{Splitthoff}
L. J. Splitthoff, J.J. Wesdorp, M. Pita-Vidal et al., \textit{Gate-tunable kinetic inductance parametric amplifier}, \href{https://journals.aps.org/prapplied/abstract/10.1103/PhysRevApplied.21.014052}{Phys. Rev. Applied 21, 014052 (2024)}.

\bibitem{Probst}
S. Probst, F. B. Song, P. A. Bushev et al., \textit{Efficient and robust analysis of complex scattering data under noise in microwave resonators}, \href{https://pubs.aip.org/aip/rsi/article/86/2/024706/360955/Efficient-and-robust-analysis-of-complex}{Rev. Sci. Instrum. 86, 024706 (2015)}.

\bibitem{Mattis}
D. C. Mattis and J. Bardeen, \textit{Theory of the Anomalous Skin Effect in Normal and Superconducting Metals}, \href{https://journals.aps.org/pr/abstract/10.1103/PhysRev.111.412}{Phys. Rev. 111, 412 (1958)}.

\bibitem{Owen}
C. S. Owen, D.J. Scalapino, \textit{Superconducting State under the Influence of External Dynamic Pair Breaking}, \href{https://journals.aps.org/prl/abstract/10.1103/PhysRevLett.28.1559}{Phys. Rev. Lett. 28, 1559 (1972)}.

\bibitem{Gao}
J. Gao, J. Zmuidzinas, A. Vayonakis, et al., \textit{Equivalence of the Effects on the Complex Conductivity of Superconductor due to Temperature Change and External Pair Breaking}, \href{https://link.springer.com/article/10.1007/s10909-007-9688-z}{J Low Temp Phys 151, 557–563 (2008)}.

\bibitem{Hu}
J. Hu, J. Matin, P. Nicaise, et al., \textit{Investigation of quasi-particle relaxation in strongly disordered superconductor resonators}, \href{https://iopscience.iop.org/article/10.1088/1361-6668/ad3f80}{Supercond. Sci. Technol. 37 055014 (2024)}.

\bibitem{Chubov}
P.N. Chubov, V. V. Eremenko and Yu. A. Plipenko, \textit{Dependence of the critical temperature and energy gap on the thickness of superconducting aluminum films}, \href{http://www.jetp.ras.ru/cgi-bin/dn/e_028_03_0389.pdf}{Sov. Phys. JETP 28, 389 (1969)}.

\bibitem{Meservey}
R. Meservey and P. M. Tedrow, \textit{Properties of Very Thin Aluminum Films}, \href{https://pubs.aip.org/aip/jap/article/42/1/51/8444/Properties-of-Very-Thin-Aluminum-Films}{J. Appl. Phys. 42, 51–53 (1971)}.

\bibitem{presto}
M. O. Tholén, R. Borgani, G. R. Di Carlo, et al., \textit{Measurement and control of a superconducting quantum processor with a fully integrated radio-frequency system on a chip}, \href{https://pubs.aip.org/aip/rsi/article/93/10/104711/2845130/Measurement-and-control-of-a-superconducting}{Rev. Sci. Instrum. 93, 104711 (2022)}.

\bibitem{Heidler}
P. Heidler, C. M. F. Scheider, K. Kustura et al., \textit{Non-Markovian Effects of Two-Level Systems in a Niobium Coaxial Resonator with a Single-Photon Lifetime of 10 milliseconds}, \href{https://journals.aps.org/prapplied/abstract/10.1103/PhysRevApplied.16.034024}{Phys. Rev. Applied 16 (2021)}.

\bibitem{Visser}
P. J. De Visser, \textit{Quasiparticle dynamics in aluminium superconducting microwave resonators}, \href{https://repository.tudelft.nl/record/uuid:eae4c9fc-f90d-4c12-a878-8428ee4adb4c}{PhD Thesis}.

\bibitem{Kaplan}
S. B. Kaplan, C. C. Chi, D. N. Langenberg et al., \textit{Quasiparticle and phonon lifetimes in superconductors}, \href{https://journals.aps.org/prb/abstract/10.1103/PhysRevB.14.4854}{Phys. Rev. B 15, 3567 (1977)}.

\bibitem{Martinis}
J. M. Martinis, M. Ansmann and J. Aumentado, \textit{Energy Decay in Superconducting Josephson-Junction Qubits from Nonequilibrium Quasiparticle Excitations}, \href{https://journals.aps.org/prl/abstract/10.1103/PhysRevLett.103.097002}{Phys. Rev. Lett. 103, 097002 (2009)}.

\bibitem{Chi}
C. C. Chi and J. Clarke, \textit{Quasiparticle branch mixing rates in superconducting aluminum}, \href{https://journals.aps.org/prb/abstract/10.1103/PhysRevB.19.4495}{Phys. Rev. B 19, 4495 (1979)}.

\bibitem{Goldie}
D. J. Goldie and S. Withington, \textit{Non-equilibrium superconductivity in quantumsensing superconducting resonators}, \href{https://iopscience.iop.org/article/10.1088/0953-2048/26/1/015004/pdf}{upercond. Sci. Technol. 26 015004 (2013)}.

\bibitem{Oppenheim}
A. V. Oppenheim, A. S. Willsky \textit{Signals \& Systems} (2nd ed.), Prentice Hall Signal Processing Series.

\bibitem{Delsing}
M. Sandberg, C. M. Wilson, F. Persson et al., \textit{Tuning the field in a microwave resonator faster than the photon lifetime}, \href{https://pubs.aip.org/aip/apl/article/92/20/203501/130983/Tuning-the-field-in-a-microwave-resonator-faster}{Appl. Phys. Lett. 92, 203501 (2008)}.

\bibitem{Wilson}
C. M. Wilson and D. E. Prober, \textit{Quasiparticle number fluctuations in superconductors}, \href{https://journals.aps.org/prb/abstract/10.1103/PhysRevB.69.094524}{Phys. Rev. B 69, 094524 (2004)}.

\bibitem{Rooij}
S. A. H. de Rooij, J. J. A. Baselmans, V. Murugesan et al.,\textit{Strong reduction of quasiparticle fluctuations in a superconductor due to decoupling of the quasiparticle number and lifetime}, \href{https://journals.aps.org/prb/abstract/10.1103/PhysRevB.104.L180506}{Phys. Rev. B 104, L180506 (2021)}.

\end{thebibliography}

\begin{thebibliography}{99}
\bibitem{s_presto}
M. O. Tholén, R. Borgani, G. R. Di Carlo, et al., \textit{Measurement and control of a superconducting quantum processor with a fully integrated radio-frequency system on a chip}, \href{https://pubs.aip.org/aip/rsi/article/93/10/104711/2845130/Measurement-and-control-of-a-superconducting}{Rev. Sci. Instrum. 93, 104711 (2022)}.

\bibitem{s_Tinkham}
M. Tinkham, "Introduction to Superconductivity" (2nd ed.), Dover Publication.

\bibitem{s_Splitthoff}
L. J. Splitthoff, J.J. Wesdorp, M. Pita-Vidal et al., \textit{Gate-tunable kinetic inductance parametric amplifier}, \href{https://journals.aps.org/prapplied/abstract/10.1103/PhysRevApplied.21.014052}{Phys. Rev. Applied 21, 014052 (2024)}.

\bibitem{s_Probst}
S. Probst, F. B. Song, P. A. Bushev et al., \textit{Efficient and robust analysis of complex scattering data under noise in microwave resonators}, \href{https://pubs.aip.org/aip/rsi/article/86/2/024706/360955/Efficient-and-robust-analysis-of-complex}{Rev. Sci. Instrum. 86, 024706 (2015)}.

\bibitem{s_Ryu}
Y. Ryu, J. Jeong, J. Suh et al, \textit{Utilizing Gate-Controlled Supercurrent for All-Metallic Tunable Superconducting Microwave Resonators}, \href{https://pubs.acs.org/doi/10.1021/acs.nanolett.3c04080}{Nano Lett. 24, 4, 1223–1230 (2024)}.

\bibitem{s_Visser}
P. J. De Visser, \textit{Quasiparticle dynamics in aluminium superconducting microwave resonators}, \href{https://repository.tudelft.nl/record/uuid:eae4c9fc-f90d-4c12-a878-8428ee4adb4c}{PhD Thesis}.

\end{thebibliography}
\end{document}